\documentclass[11pt]{article}
\usepackage[english]{babel}
\input{epsf}
\usepackage{mathtext}
\usepackage{graphicx}
\usepackage{graphics,color}
\usepackage[pdftex]{hyperref}
\newcommand{\hhh}{{\cal H}}

\newcommand{\be}{\begin{equation}}
\newcommand{\ene}{\end{equation}}
\newcommand{\ba}{\begin{array}}
\newcommand{\ea}{\end{array}}

\begin{document}

\title{Conduction band population in graphene in ultrashort strong laser field: case of massive Dirac particles}
\author{Z. Ahmadi, H. Goudarzi\footnote{h.goudarzi@urmia.ac.ir}, A. Jafari\footnote{Corresponding author, e-mail address: a.jafari@urmia.ac.ir} \\
\footnotesize\textit{Department of Physics, Faculty of Science, Urmia University, P.O.Box: 165, Urmia, Iran}}
\date{}
\maketitle

\begin{abstract}
The Dirac-like quasiparticles in honeycomb graphene lattice are taken to possess a non-zero effective mass. The charge carriers involve to interact with a femtosecond strong laser pulse. Due to the scattering time of electrons in graphene ($\tau \approx 10-100 fs$), the one femtosecond optical pulse is used to have coherence effect, and consequently, it is realized to use the time-dependent Schr$\ddot{o}$dinger equation for coupling electron with strong electromagnetic field. Generalized wavevector of relativistic electrons interacting with electric field of laser pulse leads to obtain a time-dependent electric dipole matrix element. Using the coupled differential equations of a two-state system of graphene, the density of probability of population transition between valence and conduction bands of gapped graphene is calculated. In particular, the effect of bandgap energy on dipole matrix elements in the Dirac points, and also on conduction band population is investigated. The irreversible electron dynamics is achieved when the optical pulse end. Increasing the energy gap of graphene results in a decreasing the maximum conduction band population.
\end{abstract}

\section{Introduction}

Two-dimensional honeycomb atomic lattice of graphene \cite{A,B} demonstrates a potential of great prosperity within several applications due to its peculiar electronic, optical and mechanical properties [3-6]. The electron dynamics in pristine graphene obeys massless relativistic Dirac equation, and low-energy dispersion is a linear dependence on momentum in the first Brillouin zone. There are two degenerate inequivallent valleys, located at the corners of the Brillouin zone ($K$ and $K'$ Dirac points), where the valence and conduction bands in two sublattices $A$ and $B$ related to the pseudospin cross at Dirac points. Indeed, the pristine graphene is considered as a semiconductor with zero bandgap. Actually, this is considered as a restriction for many optical and electronics applications. However inducing a bandgap in to graphene can be achieved by several methods [7-16], for instance, in graphene grown on $SiC$-substrate \cite{SZ,FV}, and, dynamically importantly, by endurance of uniaxial mechanical strain up to 22$\%$ \cite{VMP}. Thus, one can expect the massive Dirac-like electrons demonstrate a fundamentally distinct dynamical and transport behavior.

In the last few years, interaction of external electromagnetic fields with condensed matters has attracted several theoretical and experimental researches [17-23]. Thereby, a high intensity optical pulse can strongly affect the electron dynamics and, correspondingly, modify the transport and optical properties \cite{AS,MS}. The reversibility of electron dynamics in insulators in the presence of a laser fields, after the pulse end has been demonstrated \cite{VA}. Interaction of electrons in metals with a strong optical pulse gives rise to high frequency Bloch oscillation \cite{VAM}. It is observed a distinct behavior of charge carriers dynamics in graphene interacting with ultrafast (one optical oscillation) and strong ($\approx 1 V/A^{\circ}$) optical pulse comparing with both insulators and metals \cite{HK}, where the authors have show that the dynamics of quasiparticles in pristine graphene with a laser pulse is not reversible, when the laser pulse is over, and a large residual conduction band population is obtained.

In this paper, we investigate the effect of massive Dirac fermions on the interband transition in a gapped graphene interacting with an ultrashort laser pulse. We focus on the interband mixing of the valence band and the conduction band separated from each other by a $2mv^{2}_{F}$ energy gap. A schematic of Dirac point with an energy gap between conduction and valence bands is sketched in Fig. $\ref{fig1}$(a).  The wavevector of carriers is modified by the time-dependent vector potential of laser pulse, while the electric field causes to create an electric dipole moment between the states of the conduction and valence bands. Influence of an energy gap between conduction and valence bands can significantly affect mentioned dipole moments, since the excitation of electrons may be a nonzero value at the Dirac points. This paper is organized as follows. In Sec. 2 we present the proposed structure and related formalism to obtain the explicit form of dipole matrix elements and exact relation for time-dependent conduction band population (CBP). The numerical results and discussion about the influence of energy gap between conduction and valence band of graphene on the dynamics of quasiparticles are presented in the section 3. Finally, we summarize our findings in Sec. 4.

\section{Model and formalism}

\begin{figure}[ht]
\centering
\includegraphics[scale=0.6]{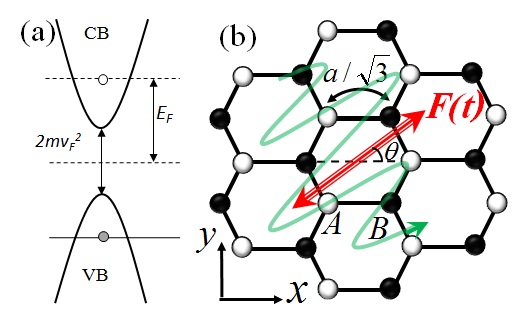}
\caption{(a),(b)}
\label{fig1}
\end{figure}

In order to study the effect of population transition between valence and conduction bands in gapped graphene illuminated by ultrafast laser field we consider an optical laser pulse that is normally incident on a graphene sheet with linear polarization in graphene plane, as shown in Fig. $\ref{fig1}$(b). The gapped graphene Hamiltonian $H_{0}$ is introduced within the nearest-neighbor tight-binding model \cite{TPA} by a $2\times 2$ matrix of form: 
\begin{equation}
H_{0}=\left(\begin{array}{cc}
\alpha-E_{F}&\gamma f(\textbf{k})\\
\gamma f^{\ast}(\textbf{k})&-\alpha-E_{F}
\end{array}\right),
\end{equation}
where $\gamma=-3.03 \: eV$ and $E_{F}$ are the hopping integral and Fermi energy, respectively. $\alpha=m\sigma_{z}$ denotes bandgap of graphene, which $m$ is effective mass of Dirac fermions, $\sigma_{z}$ is the Pauli matrix. The off-diagonal term of Hamiltonian is given by
\begin{equation}
f(\textbf{k})=\exp\left(\frac{iak_{x}}{\sqrt{3}}\right)+2\exp\left(\frac{-iak_{x}}{2\sqrt{3}}\right)\cos\left(\frac{ak_{y}}{2}\right)=\left|f(\textbf{k})\right|e^{i\varphi_{k}},
\end{equation}
where $a=2.46 \: A^{\circ}$ is lattice constant and $\textbf{k}$ denotes the wavevector of electron. The energy spectrum of Hamiltonian $H_{0}$ consist of conduction band $(\pi^{\ast})$ and the valence bands $(\pi)$ with the energy dispersion $\epsilon_{c,v}=\pm\sqrt{m^{2}+\gamma^{2}\left|f(\textbf{k})\right|^{2}}-E_{F}$. By solving the Dirac equation for Hamiltonian Eq. (1), the corresponding wavefunctions are given as
\begin{equation}
\psi^{(c)}_{\textbf{k}}(\textbf{r})=\frac{e^{i\textbf{k}\cdot\textbf{r}}}{\sqrt{2}}\left(\begin{array}{cc}
\sqrt{\frac{\epsilon_{c}+E_{F}+m}{\epsilon_{c}+E_F}}e^{i\varphi_{k}} \\
\sqrt{\frac{\epsilon_{c}+E_{F}-m}{\epsilon_{c}+E_F}}
\end{array}\right),
\end{equation}
\begin{equation}
\psi^{(v)}_{\textbf{k}}(\textbf{r})=\frac{e^{i\textbf{k}\cdot\textbf{r}}}{\sqrt{2}}\left(\begin{array}{cc}
-\sqrt{\frac{\epsilon_{c}+E_{F}-m}{\epsilon_{c}+E_F}}e^{i\varphi_{k}} \\
\sqrt{\frac{\epsilon_{c}+E_{F}+m}{\epsilon_{c}+E_F}}
\end{array}\right),
\end{equation}
where two components of wavefunctions belong to sublattices $A$ and $B$, respectively, according Fig. $\ref{fig1}$(a). Opening a bandgap in graphene does not change the linearity of energy dispersion near the Dirac points, and the band structure of doping graphene is similar to the band structure of pristine graphene. However, existence of bandgap can be an important feature for electronics and optical applications, and actually, gives rise to another behavior of relativistic quasiparticles in transitions between distinct bands. Normally incident optical pulse on a gapped graphene sheet is taken to be:
\begin{equation}
F(t)=F_{0}e^{-u^{2}}(1-2u^{2}).
\end{equation}
This form of optical pulse is an idealization of the actual 1.5 oscillation pulses used in recent experiments \cite{AS,MS}. $F_{0}$ is amplitude of pulse, $u=t/\tau$, and $\tau$ denotes the pulse lenght, which is set $\tau=1 \: fs$ corresponding to carrier frequency $\omega=1.5 \: eV/\hbar$. We suppose the pulse is linearly polarized, so that the plane of polarization is charactrized by angle $\theta$ measured relative to $x$-axis (see Fig. $\ref{fig1}$(b)). The electron dynamics in gapped graphene may interact with time-dependent electric field of optical pulse, in which its hamiltonian reads:
\begin{equation}
\hhh (t)=H_{0}+e\textbf{F}(t)\cdot\textbf{r},
\end{equation}
where $\textbf{r}$ is a two-dimensional vector. If length of pulse is less than the characteristic electron scattering time, which is $10-100 \: fs$ [30-35], then the electron dynamics in external electric field may be coherent, and consequently, is described by time-dependent Schr$\ddot{o}$dinger equation.

The electric field of optical pulse accelerates the electrons of graphene in the direction of the electric field polarization through the graphene plane, and also changes the wave vector $\textbf{k}$ of electrons. Moreover, the electric field can affect both intraband and interband electron dynamics. The interband electron dynamics causes coupling the conduction and valence band states, and makes the redistribution of electrons between two bands. First, we describe the electron dynamics within a single band in the reciprocal space, which is determined by acceleration theorem
\begin{equation}
\hbar\frac{d\textbf{k}(t)}{dt}=e\textbf{F}(t).
\end{equation}
For an electron with initial momentum $\textbf{q}$ at the first Brillouin zone in corners of honeycomb lattice, one can obtain the generalized time-dependent wavevector $\textbf{k}(\textbf{q},t)$
\begin{equation}
\textbf{k}(\textbf{q},t)=\textbf{q}+\frac{e}{\hbar}\int^{t}_{-\infty}\textbf{F}(t_{1})dt_{1},
\end{equation}
and corresponding wavefunction can be expressed as Houston functions \cite{WH}, 
\begin{equation}
\Phi^{(H)}_{\delta\textbf{q}}(\textbf{r},t)=\psi^{(\delta)}_{\textbf{k}(\textbf{q},t)}(\textbf{r})\exp\left(\frac{-i}{\hbar}\int^{t}_{-\infty}dt_{1}E_{\delta}\left[\textbf{k}(\textbf{q},t_{1})\right]\right),
\end{equation}
where $\psi^{(\delta)}_{\textbf{k}(\textbf{q},t)}(\textbf{r})$ is the Dirac spinors of Eqs. (3) and (4). Finally, solving of Schr$\ddot{o}$dinger equation for interacted Hamiltonian $\hhh (t)$ is given by the superposition of Houston functions:
\begin{equation}
\Psi_{\textbf{q}}(\textbf{r},t)=\sum_{\delta=c,v}\beta_{\delta\textbf{q}}(t)\Phi^{(H)}_{\delta\textbf{q}}(\textbf{r},t),
\end{equation}
where $\beta_{\delta\textbf{q}}(t)$ is corresponding time-dependent expansion coefficients. Due to acceleration theorem, the electrons which belong to different bands but have the same initial wavevector $\textbf{q}$ possesses the same wave vector $\textbf{k}(\textbf{q},t)$ at later moment of time $t$. Coupling of conduction and valence band states in external electric field is determined by interband dipole matrix element. The interband dipole matrix element is diagonal in reciprocal space, so the states with different wave vector are not coupled by the pulse field. Such coupling of states with equal value of $\textbf{q}$ is the property of coherent dynamics. Substituting wavefunctions expressed in Eq. (10) into the time-dependent Schr$\ddot{o}$dinger equation results in two coupled differential equations
$$
\frac{d\beta_{c\textbf{q}}(t)}{dt}=\frac{-i}{\hbar}\textbf{F}(t)\cdot\textbf{Q}_{\textbf{q}}(t)\beta_{v\textbf{q}}(t),
$$
\begin{equation}
\frac{d\beta_{v\textbf{q}}(t)}{dt}=\frac{-i}{\hbar}\textbf{F}(t)\cdot\textbf{Q}^{\ast}_{\textbf{q}}(t)\beta_{c\textbf{q}}(t),
\end{equation}
in which one can easily obtain the interband dipole matrix element $\textbf{Q}_{\textbf{q}}(t)$ as
\begin{equation}
\textbf{Q}_{\textbf{q}}(t)=\textbf{D}\left[\textbf{k}(\textbf{q},t)\right]\exp\left\{\frac{-i}{\hbar}\int^{t}_{-\infty}dt_{1}\left[\epsilon_{c}\left[\textbf{k}(\textbf{q},t_{1})\right]-\epsilon_{v}\left[\textbf{k}(\textbf{q},t_{1})\right]\right]\right\},
\end{equation}
where $\textbf{D}(\textbf{k})=\left\langle\psi^{(c)}_\textbf{k}\left|e\textbf{r}\right|\psi^{(v)}_\textbf{k}\right\rangle$ is the dipole matrix element between distinct bands with wavevector $\textbf{k}$. 
We can obtain the following expressions for the interband dipole matrix elements by substituting Eqs. (3) and (4) into expression of $\textbf{D}(\textbf{k})$, as follows:
\begin{equation}
D_{j}(\textbf{k})=\sqrt{1-\left(\frac{m}{\epsilon_{c}+E_{F}}\right)^2} Z_j-\frac{ie}{2}\frac{m\partial\epsilon_{c}/\partial k_{j}}{(\epsilon_{c}+E_F)\sqrt{(\epsilon_{c}+E_{F})^{2}-m^{2}}} \: ;\ \ \ (j=x,y),
\end{equation}
where we have
$$
Z_{x}=\frac{ea}{2\sqrt{3}}\frac{1+\cos(\frac{ak_{y}}{2})\left[\cos(\frac{3ak_{x}}{2\sqrt{3}})-2\cos(\frac{ak_{y}}{2})\right]}{1+4\cos(\frac{ak_{y}}{2})\left[\cos(\frac{3ak_{x}}{2\sqrt{3}})+\cos(\frac{ak_{y}}{2})\right]},
$$
$$
Z_{y}=\frac{ea}{2}\frac{\sin(\frac{ak_{y}}{2})\sin(\frac{3ak_{x}}{2\sqrt{3}})}{1+4\cos(\frac{ak_{y}}{2})\left[\cos(\frac{3ak_{x}}{2\sqrt{3}})+\cos(\frac{ak_{y}}{2})\right]}.
$$
If the energy gap is taken to be zero, then dipole matrix element reduces to that has been obtained in Ref. \cite{HK}. For solving the coupled differential equations of two-state system Eq. (11), which describes the interband electron dynamics and, indeed, determines the mixing of the states in the presence of electric field of laser pulse, we consider the initial condition as $(\beta_{v\textbf{q}},\beta_{c\textbf{q}})=(1,0)$, that it means just before interaction between electric field of the pulse and quasiparticle in gapped graphene, all states of valence bands are occupied, and consequently, all states of conduction bands are empty.
The mixing of the states of different bands and the dynamics of an electron which is initially in valence band characterized by time-dependent expansion coefficients $\left|\beta_{c\textbf{q}}(t)\right|^{2}$. The time-dependent total transition of conduction band is expressed by the following expression:
\begin{equation}
N_{CB}(t)=\sum_{\textbf{q}}\left|\beta_{c\textbf{q}}(t)\right|^{2},
\end{equation}
where the sum is over the all momentum in the first Brillouin zone. 

\section{Numerical results and discussion}

In this section, we discuss, in detailed, about the conduction band population (CBP), resulting from Eq. (14) with respect to bandgap energy of graphene. Very recently in Ref. \cite{HK}, the authors have stated that the gapless energy dispersion of pristine graphene is responsible to the irreversible electron dynamics in an optical field. Regarding the fact that it is possible to implement an energy gap in graphene, it can be realized to show how nonzero effective mass of relativistic quasiparticles influence the electron dynamics in gapped graphene interacted with a ultrashort laser field. According to Eq. (13), the interband dipole matrix elements $D_x(\textbf{k})$ and $D_y(\textbf{k})$ are found to be a complex functions, and strongly depend on the electron wave vector $\textbf{k}$. Their absolute value is singular at the Dirac points, $K$ and $K'$ in the corners of Brillouin zone, as shown in Fig. $\ref{fig3}$(a) and (b), where the energy gap causes to fall strongly the height of sharp peaks. There is no coupling at the center of the Brillouin zone ($\Gamma$ point). In Fig. $\ref{fig4}$, we plot the total CBP $N_{CB}(t)$ as a function of time for various magnitudes of energy gap. Firstly, in agreeing with gapless graphene, we obtain the irreversible feature of electron dynamics in strong electric field of a optical pulse when the pulse is over, and secondly, the residual of CBP is considerable even in high energy gaps. However, the energy gap in graphene causes to decrease a little the maximum of the CBP. To more clarify, the dependence of maximum and residual CBP on the massive Dirac-like electrons is particularly presented in Fig. $\ref{fig5}$. We observe that, for low energy gaps, the maximum and residual CBP show fundamentally distinct behaviors, whereas they are equal and, of course, decrease rapidly for large energy gaps.

Further, we consider the transition probability (TP) $|\beta_{CB}(t)|^2$ for four Dirac points (see Fig. $\ref{fig6}$) of Brillouin zone in terms of different energy gaps, for $m=0.1$ in Figs. $\ref{fig7}$ and $m=0.5$ in Figs. $\ref{fig8}$, respectively. We find unit probability with oscillating function of time in Dirac points for zero energy gap, which is not shown here. By applying an energy gap, the TP decreases, and sharp peaks of probability oscillating are found. Increasing energy gap causes to more decreasing the TP. This feature is shown in figure $\ref{fig8}$. For right side of Dirac points in Fig. $\ref{fig6}$, the TP occurs for negative values of electric field of optical pulse, and for left side once it happens for positive electric field.

Using the acceleration theorem, we investigate the distribution of electrons around Dirac points $K$ and $K'$ during the pulse. In Figs. $\ref{fig9}$, we present nonuniform distribution of the CBP versus wave vectors in the reciprocal space for nonzero energy gap and field strength $F_0=1\: V/A^\circ$ and polarization angle $\theta=0$ in different times. Regarding that, the maximum peak of laser field is at $t=0$, for $t\leq -0.75 \: fs$ the field is negative which accelerates the electrons to the right. Then the field changes its sign, the electrons start to move left. After the pulse end, the distribution becomes completely symmetric, and interference fringes at the $k_x\approx 1 \: A^{\circ -1}$ and $-1 \: A^{\circ -1}$ Dirac points. We actually have considerable hot points in the reciprocal space for any magnitude of mass term. By increasing the gap, the area of hot points reduces due to decreasing the TP. The symmetric conduction band redistribution is found in Dirac points when the pulse end. 

\section{Conclusion}

In summary, in this paper we have studied transition of electrons from valence band to conduction band in graphene with nonzero bandgap (in fact, more real graphene). The relativistic massive Dirac electrons interact with ultrashort (one optical oscillation) and strong ($\approx 1 V/A^{\circ}$) optical pulse. The electron dynamics of system was considered to be in coherent case due to no appearance of electron scattering. The effect of bandgap on the energy dispersion, and correspondingly in Fermi wave vectors of graphene has been considered in the transferred conduction band population and transition probability. Dipole matrix element of system has been found to be a complex function in $x$- and $y$-components. The numerical calculations of total transition rate for all points of first Brillouin zone in reciprocal space have shown weak dependence of CBP on the bandgap parameter for $0\leq m\leq 0.6$. Due to singularity of dipole matrix element in reciprocal space, existence of a gap in graphene results in irreversible electron dynamics and nonuniform distribution of the CBP, which is in contrast with isolators.


\newpage

\textbf{Figure captions}\\
\textbf{Figure 1.} (Color online)(a) Opening bandgap $E_g=2mv^{2}_{F}$ in energy dispersion of graphene near the Dirac points in reciprocal space and level of Fermi energy $E_F$, VB and CB correspond to valence and conduction band, respectively, and (b) sketch of lattice structure of graphene which is illuminated by laser pulse. $A$ and $B$ correspond to two inequivalent sublattices. Optical pulse is perpendicularly glinted to the graphene sheet and its electric field makes angle $\theta$ with $x$-direction in graphene plane.\\
\textbf{Figure 2.}(a), (b) (Color online) Plot of absolute of $x$-component of dipole matrix element as a function of wave vectors, (a) for energy gap $m=0.5$ and (b) $m=0.3$.\\
\textbf{Figure 3.} (Color online) Population of the conduction band $N_{CB}(t)$ as a function of time for different values of energy gap, which for all states we take $F_{0}=1 V/A^{\circ}$ and $\theta=0$.\\
\textbf{Figure 4.} (Color online) Curves of maximum and residual conduction band population as a function of bandgap.\\
\textbf{Figure 5.} (Color online) Dirac points in the reciprocal space in the first Brillouin zone.\\
\textbf{Figure 6.}(a),(b),(c) and (d) (Color online) Plot of transition probability versus time for four Dirac $K$ and $K'$ points as shown in Fig. 5, for energy gap $m=0.1$.\\
\textbf{Figure 7.}(a),(b),(c) and (d) (Color online) Plot of transition probability versus time for four Dirac $K$ and $K'$ points as shown in Fig. 5, for energy gap $m=0.5$.\\
\textbf{Figure 8.}(a),(b),(c) and (d) (Color online) Conduction band distribution of electrons at six Dirac points as a function of wave vector $\textbf{k}$ for energy gap $m=0.2$ in various times of optical pulse. Only the first Brillouin zone of the reciprocal space is shown, $F_{0}=1 V/A^{\circ}$ and $\theta=0$. (a) the time is $t=-0.75 \;fs$, (b) $t=0$, (c) for $t=0.75 \;fs$ and (d) in $t=2.25 \;fs$. 

\newpage

\begin{figure}[ht]
\centering
\includegraphics[scale=0.4]{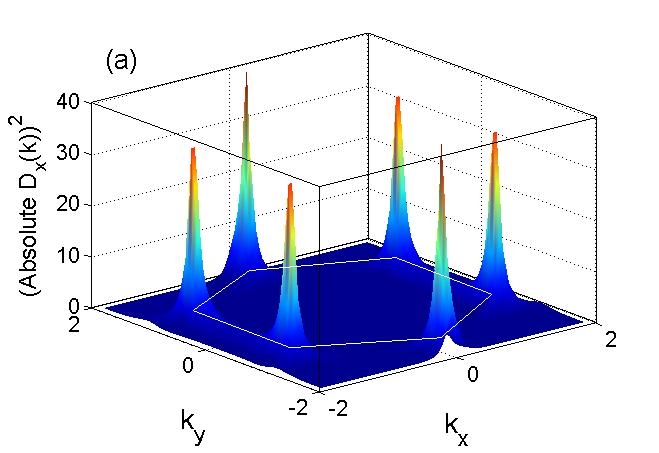}
\end{figure}

\begin{figure}[ht]
\centering
\includegraphics[scale=0.4]{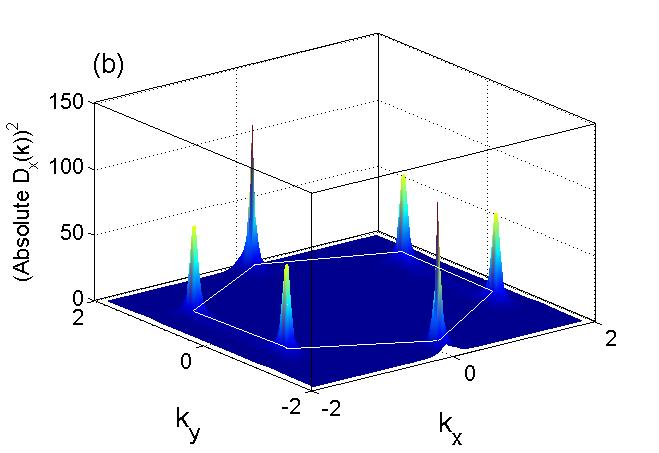}
\caption{(a), (b)}
\label{fig3}
\end{figure}

\begin{figure}[ht]
\centering
\includegraphics[scale=0.4]{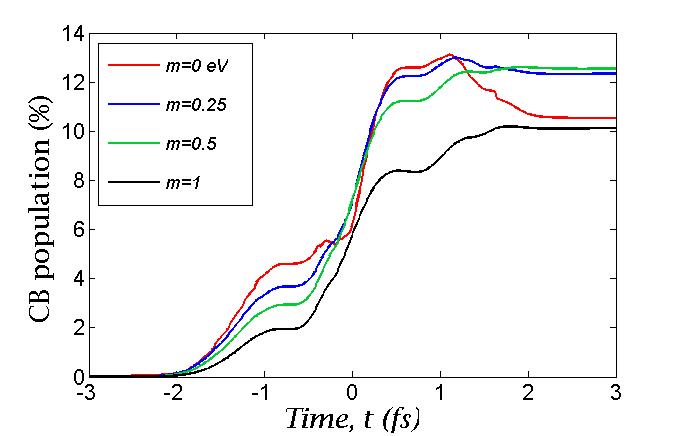}
\caption{}
\label{fig4}
\end{figure}

\begin{figure}[ht]
\centering
\includegraphics[scale=0.4]{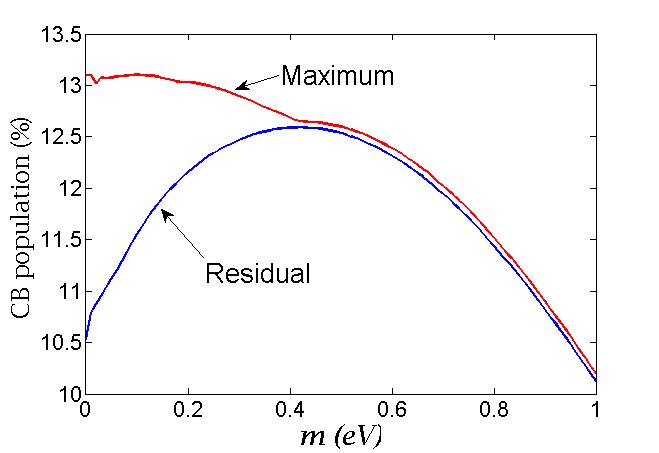}
\caption{}
\label{fig5}
\end{figure}

\begin{figure}[ht]
\centering
\includegraphics[scale=1]{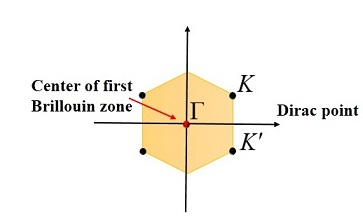}
\caption{}
\label{fig6}
\end{figure}

\begin{figure}[ht]
\centering
\includegraphics[scale=0.4]{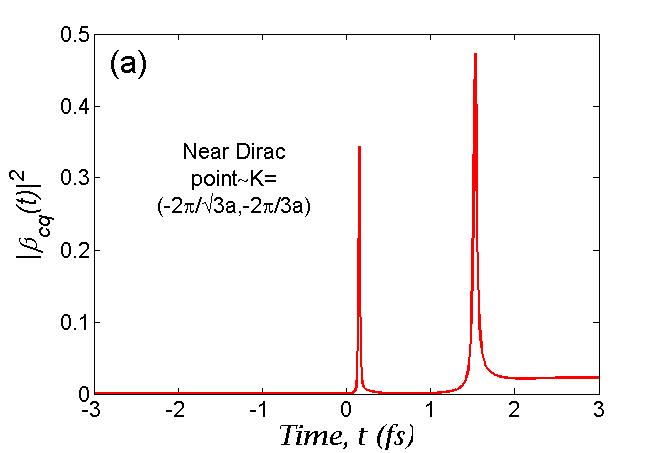}
\end{figure}

\begin{figure}[ht]
\centering
\includegraphics[scale=0.4]{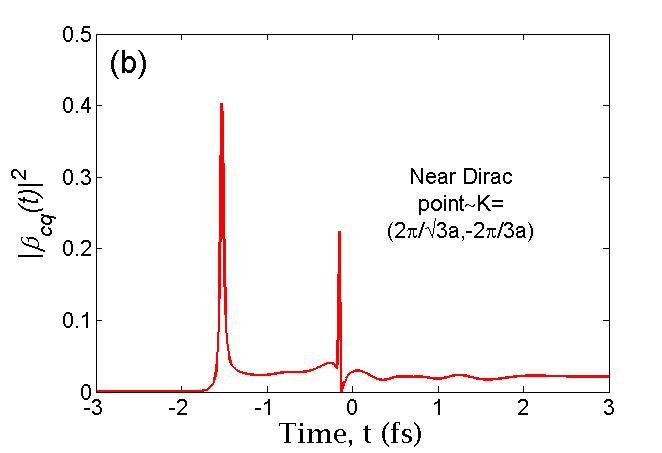}
\end{figure}

\begin{figure}[ht]
\centering
\includegraphics[scale=0.4]{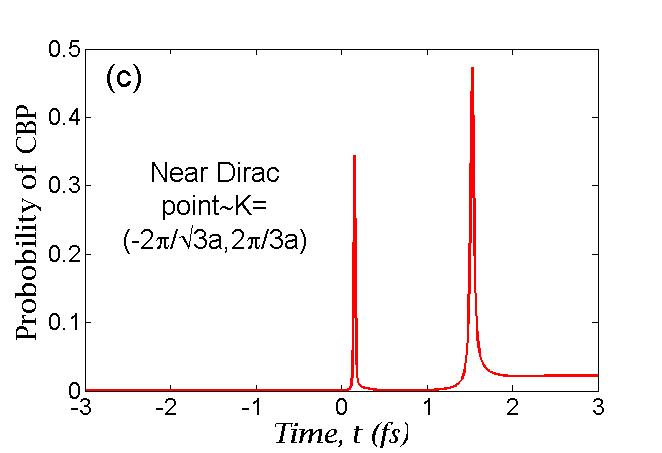}
\end{figure}

\begin{figure}[ht]
\centering
\includegraphics[scale=0.4]{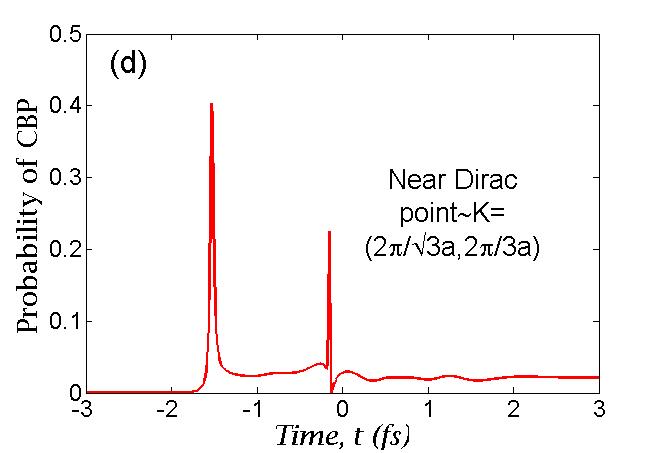}
\caption{(a), (b), (c), (d)}
\label{fig7}
\end{figure}

\begin{figure}[ht]
\centering
\includegraphics[scale=0.4]{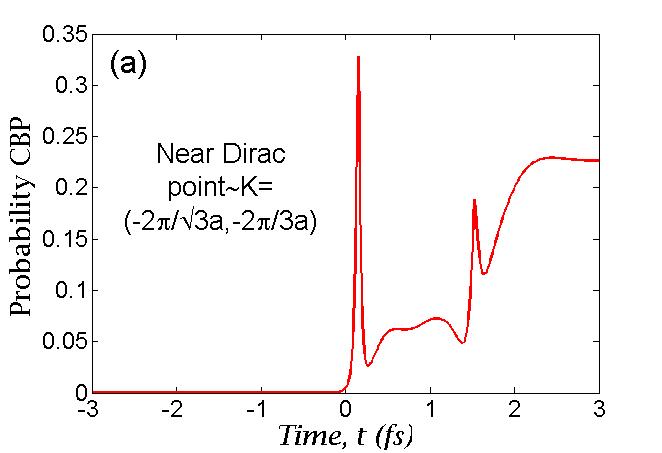}
\end{figure}

\begin{figure}[ht]
\centering
\includegraphics[scale=0.4]{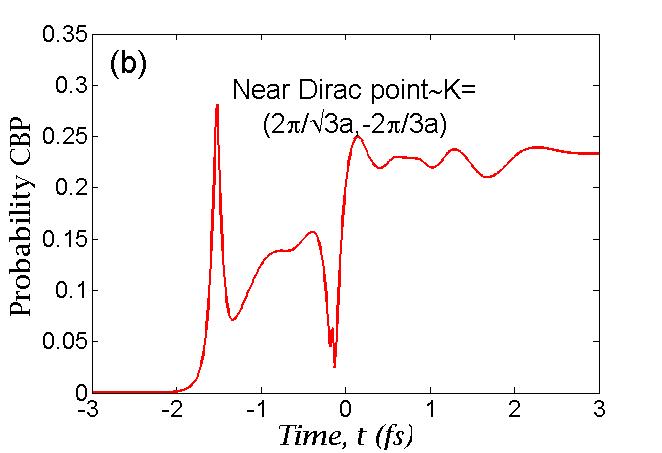}
\end{figure}

\begin{figure}[ht]
\centering
\includegraphics[scale=0.4]{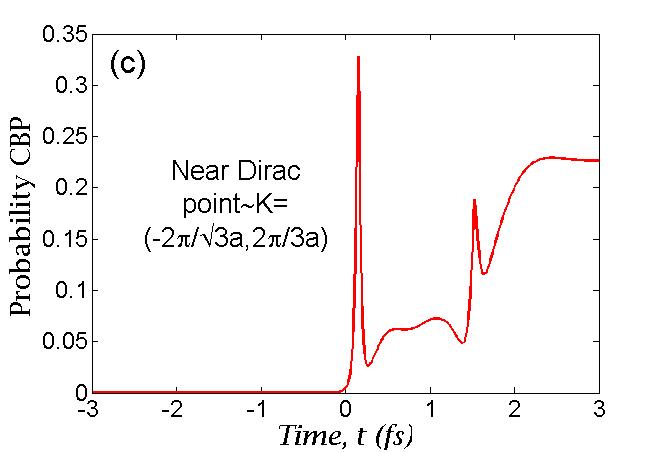}
\end{figure}

\begin{figure}[ht]
\centering
\includegraphics[scale=0.4]{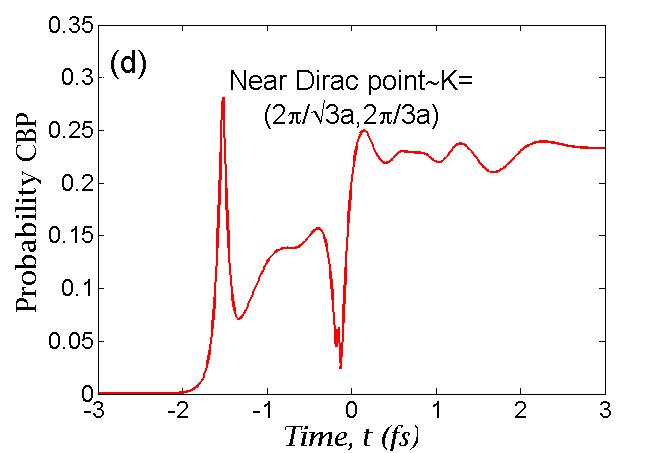}
\caption{(a), (b), (c), (d)}
\label{fig8}
\end{figure}

\begin{figure}[ht]
\centering
\includegraphics[scale=0.3]{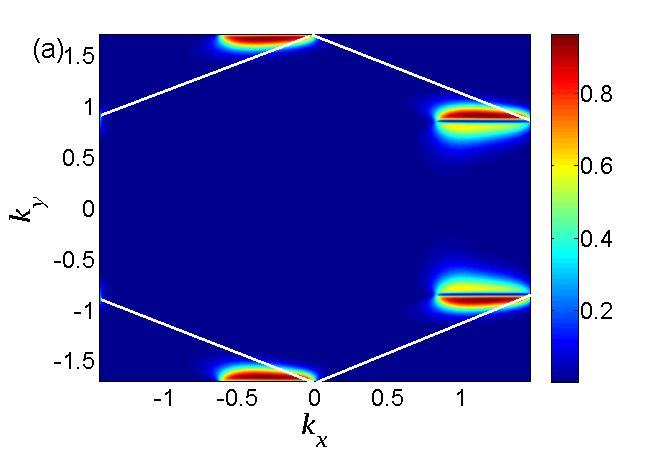}
\end{figure}

\begin{figure}[ht]
\centering
\includegraphics[scale=0.3]{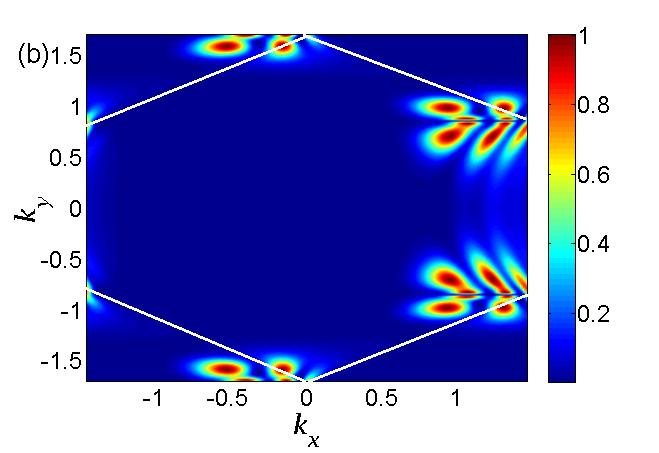}
\end{figure}

\begin{figure}[ht]
\centering
\includegraphics[scale=0.3]{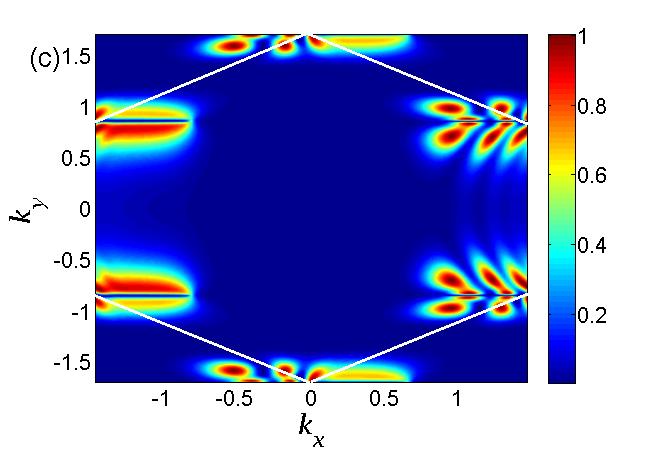}
\end{figure}

\begin{figure}[ht]
\centering
\includegraphics[scale=0.3]{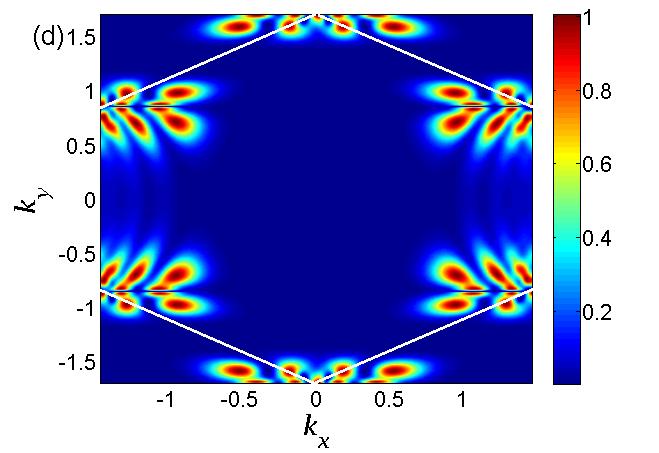}
\caption{(a), (b), (c), (d)}
\label{fig9}
\end{figure}

\end{document}